\documentclass{article}

\usepackage{PRIMEarxiv}
\usepackage[dvipsnames]{xcolor}
\usepackage[utf8]{inputenc} 
\usepackage[T1]{fontenc}    
\usepackage{hyperref}       
\usepackage{url}            
\usepackage{booktabs}       
\usepackage{colortbl}
\usepackage{multicol}
\usepackage{tabularray}
\usepackage{nicefrac}       
\usepackage{microtype}      
\usepackage{lipsum}
\usepackage{fancyhdr}       
\usepackage{graphicx}       
\graphicspath{{media/}}     

\author{
  Shushanta Pudasaini \\
  Technological University Dublin \\
  \texttt{D23129142@mytudublin.ie} \\
   \And
  Luis Miralles-Pechuán\\
  Technological University Dublin\\
  \texttt{luis.miralles@TUDublin.ie} \\
  \AND
  David Lillis \\
  University College Dublin \\
  \texttt{david.lillis@ucd.ie} \\
  \And
  Marisa Llorens Salvador \\
  Technological University Dublin\\
  \texttt{marisa.llorens@TUDublin.ie} \\
}

\begin{document}

\title{Survey on Plagiarism Detection in Large Language Models: The Impact of ChatGPT and Gemini on Academic Integrity}

\maketitle

\begin{abstract}
The rise of Large Language Models (LLMs) such as ChatGPT and Gemini has posed new challenges for the academic community. With the help of these models, students can easily complete their assignments and exams, while educators struggle to detect AI-generated content. This has led to a surge in academic misconduct, as students present work generated by LLMs as their own, without putting in the effort required for learning. As AI tools become more advanced and produce increasingly human-like text, detecting such content becomes more challenging. This development has significantly impacted the academic world, where many educators are finding it difficult to adapt their assessment methods to this challenge.

This research first demonstrates how LLMs have increased academic dishonesty, and then reviews state-of-the-art solutions for academic plagiarism in detail. A survey of datasets, algorithms, tools, and evasion strategies for plagiarism detection has been conducted, focusing on how LLMs and AI-generated content (AIGC) detection have affected this area. The survey aims to identify the gaps in existing solutions. 
Lastly, potential long-term solutions are presented to address the issue of academic plagiarism using LLMs based on AI tools and educational approaches in an ever-changing world.

\end{abstract}



\section{Introduction}
1 out of 10 assignments was detected using AI tools from more than 200 million writing assignments reviewed by the AI detector engine of Turnitin over the past year \cite{edweekDataReveal}. Specifically, the introduction of ChatGPT, a revolutionary LLM-based conversational engine, has significantly changed various industries and domains \cite{kalla2023study}, including academia. The capabilities of highly advanced LLMs have affected the academic world in various ways. For example, students are using ChatGPT to complete their homework assignments \cite{tossell2024student} and to pass challenging exams like the Graduate Record Examination (GRE) and Scholastic Assessment Test (SAT) \cite{businessinsiderGPT4Bar}. This has raised concerns about the current evaluation systems used in academic institutions. Lecturers and universities are struggling to detect fraudulent activities by students \cite{theguardianMakesPlagiarism}, with plagiarism being one of the major issues. In the past, plagiarism was mostly done by presenting a document which included paragraphs from other sources without being referred to, but with the emergence of LLMs, students can now use LLMs to generate text and complete their assignments entirely. This act of using the text generated by LLMs and claiming their own work is referred to here as AI-generated plagiarism. This research explores the impact of LLMs on plagiarism and mainly focuses on detecting AI-generated plagiarism. 

In a survey by Intelligent.com conducted in May 2023 among 3,017 high school and college students, it was found that nearly one-third of the students admitted to using ChatGPT to complete their homework \cite{forbesEducatorsBattle}. The dependency of students on such tools leads to the loss of creativity and learning ability \cite{smolansky2023educator}. Due to such threats, several universities have decided to ban the usage of ChatGPT \cite{businessinsiderHereSchools}. Apart from students, researchers have also been misusing this technology. Anyone with little knowledge and research experience can use ChatGPT to write academic content.  A recent publication in Nature states that at least four preprints were submitted to their journal, including ChatGPT as a co-author  \cite{natureChatGPTListed}. The disruption created by this phenomenon in the field of scientific publishing has been such that journals have strictly banned listing ChatGPT as a co-author \cite{theguardianScienceJournals} and set out new authorship guidelines for AI-generated text \cite{nihScienceJournals}. Due to such problems, LLMs have been a major issue in academia.

LLMs, or language models, are deep learning-based models designed for various Natural Language Processing (NLP) tasks. 
They offer remarkable capabilities such as brainstorming, generating counterarguments, creating summaries and abstracts, and correcting grammar. LLMs like ChatGPT can rephrase text and produce text almost indistinguishable from human writing\cite{ChatGPTparaphraser}. These models can handle simple tasks such as generating an essay on a given topic as well as complex tasks like writing a research paper on a challenging problem \cite{natureChatGPTListed}. Rehan Haque, Founding Director of Metatalent.ai, suggests that we have reached a point where entire programming projects can be done using AI algorithms such as LLMs and use another AI tool to modify to make AI undetectable \cite{dailymailHalfStudents}. Thus, the rise of such generative AI tools and their capacity to generate even more human-like text poses a significant threat to academic integrity \cite{ChatGPT_acedmic_threat}.

After the release of ChatGPT on November 30, 2022, research has been conducted on both sides: developing more intelligent LLMs and models that can detect such AI-generated content. The task of detecting whether a text is generated by AI-based algorithms or humans is termed Artificial Intelligence Generated Content (AIGC) detection \cite{arxivArguGPTEvaluating}. From the 94 million training parameter ELMO model in 2019 to the recent 1.76 trillion parameter GPT-4 model, the evolution of the LLMs and their capabilities have been growing daily \cite{xi2023rise}. With the projected rapid development of highly capable LLMs, the quality of outputs is increasing, making it more difficult to detect \cite{zhao2023survey}.

Different approaches, such as training classifiers, watermarking, and zero-shot approaches, have been developed to detect text generated from LLMs.  Based on these approaches, algorithms such as DetectGPT  \cite{arxivDetectGPTZeroShot}, RADAR  \cite{arxivRADARRobust}, Ghostbuster \cite{arxivGhostbusterDetecting}, GPT-Sentinel \cite{arxivGPTSentinelDistinguishing} amongst others are being developed to identify AI-generated content. OpenAI, the creator of ChatGPT, introduced its AIGC detection tool two months after its release. However, OpenAI states that the detector is not fully reliable \cite{openaiClassifierIndicating}. Similarly, several AIGC detector tools and software such as CopyLeaks, Turnitin, GPTZero, and Crossplag have been released for the general use of the public to identify AI-generated content. On the other hand, different techniques to attack or evade such AIGC detectors have also been developed and are an active area of research \cite{cai2023evade}. Evasion techniques such as prompting \cite{arxivLargeLanguage}, recursive paraphrasing  \cite{arxivParaphrasingEvades}, authorship obfuscation \cite{arxivAuthorshipObfuscation}, and sentence or word substitution have been developed to point out the failures in the AIGC detector tools. 

The main contribution of this research is the comprehensive survey of existing algorithms, tools, datasets, and evasion strategies developed for addressing academic misconduct, particularly in plagiarism detection and AIGC detection. Through a simple experiment, we evaluated the reliability of existing tools in detecting misconduct. Based on the experiment and survey, we discussed the reliability and feasibility of addressing the problem. In addition, various alternative educational solutions to address the issue have also been discussed. 

This paper is organized as follows. Section \ref{PlagiairismProblem} identifies different problems due to academic dishonesty with the emergence of LLMs and demonstrates how such LLMs have affected academia by presenting different ways such tools are improvising academic cheating. A survey of existing algorithms, datasets, and tools for detecting academic cheating methods such as plagiarism and generating text with AI is performed in this section \ref{existingsolutions}. Different evasion techniques have also been discussed in Section \ref{existingsolutions}. Based on the detection algorithms and evasion strategies survey, the limitations and gaps observed in current solutions have been discussed in Section \ref{gaps}. Finally, the feasibility of technical solutions for AIGC detection and other alternative educational solutions have been discussed in Section \ref{discussion}
 
\section{The plagiarism problem in Academia} \label{PlagiairismProblem}
This section discusses the impact of LLMs in academia, how LLMs have been utilized for academic misconduct, and the proliferation of LLMs, which has resulted in increasingly human-like text that poses a serious threat to academic integrity.

\subsection{Rise of Large Language Models (LLMs)}
 In recent years, deep learning models have been capable of generating different types of data. For example, DALL-E 3 from OpenAI can generate images, while models like ChatGPT, Gemini, and Perplexity can generate text. AudioGen and MusicGen from MetaAI can generate audio, and GPT-4 from OpenAI can generate multimodal data, meaning it can handle different types of data. These models require a simple, optimized prompt so that users can get the desired data. This domain of AI, referred to as Generative AI, is growing exponentially. The market size of generative AI was reported to be 44 billion USD in 2023 and is projected to reach 66.62 billion USD in 2024 \cite{statistaGenerativeWorldwide}. OpenAI, one of the leading companies in this field, reported that 100 million use ChatGPT every week and over 2 million developers are building generative AI applications from the API service provided by the company \cite{thevergeChatGPTContinues}. 

 The content generated from Generative AI tools is AI Generated Content (AIGC). With the increasing use of Generative AI tools, a significant amount of AIGC has been detected on the internet, with some estimates suggesting that 10\% of internet content is already AI-generated \cite{zrealityPercentInternet}. AIGC presents various threats and challenges, including ethical concerns, harmful or inappropriate content, bias, over-reliance, misuse, digital divide, academic misconduct, security, and privacy \cite{threats_IGC}. This research mainly focuses on AI-generated textual content.

Language Modelling(LM) began with
statistical learning methods such as building word prediction models called N-gram language models based on Markov assumption in the 1990s \cite{ieeeErgodicHidden}.
However, these N-gram language models suffered from issues like increased computational complexity and overfitting because of high dimensional data, also called the curse of dimensionality. The focus then shifted towards Neural Language Models (NLMs), which utilized deep learning architectures like multi-layer perceptron (MLPs) \cite{springerHybridApproach} 
and recurrent neural networks (RNNs) \cite{arxiv_rnn} to characterize the probability of word sequences. Subsequently, Pre-trained Language Models (PLMs) such as ELMO \cite{arxivDeepContextualized} and BERT \cite{paperswithcodePapersWith} were introduced, which could capture the context-aware representation of any text. LLMs have been developed based on pre-existing PLMs, which have completely transformed the landscape of text generation. These LLMs are now readily available to the public through products from tech giants like ChatGPT and Gemini. They enable remarkable text generation, allowing individuals to generate ideas, create academic content, summarize large amounts of text, generate code, paraphrase any text, and more. However, this widespread accessibility has raised serious ethical concerns regarding its use \cite{huallpa2023exploring}. 

\begin{figure}[hbt!]
    \centering
    \includegraphics[width=0.65\textwidth,height=\textheight,keepaspectratio]{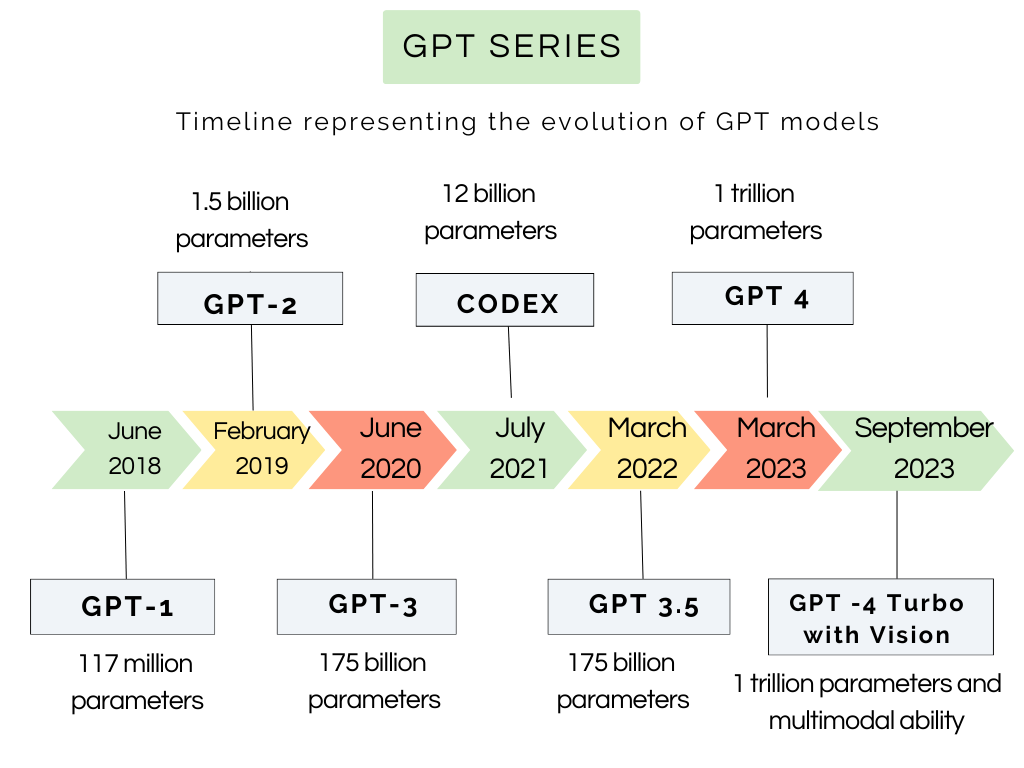}
    \caption{Timeline indicating the release date and parameter of different GPT models by OpenAI.
}
    \label{fig: GPT models timeline}
\end{figure}

The growth of LLMs has been reshaping the AI landscape due to their rapid growth in size and capabilities at a staggering pace \cite{huggingfaceLargeLanguage}.  This is evident in the timeline of various GPT models released by OpenAI in recent years, as depicted in Figure \ref{fig: GPT models timeline}. 

Apart from OpenAI, many tech companies like Microsoft and Google are involved in the LLMs development race. Open-source platforms like HuggingFace are also in the race to bring out more powerful and intelligent LLMs. Currently, HuggingFace has 535,131 models, over 250,000 datasets, and more than 250,000 spaces uploaded on their platform \cite{originalityHuggingFaceStatistics}. This indicates that LLMs will continue to improve over time and can generate more human-like text in the future, making AIGC detection more challenging.

\subsection{Academic Misconduct in the LLMs Era}
According to a survey in the USA, it was found that 82 \% of undergraduate students admitted to some form of misconduct while submitting their assignments \cite{jstorAcademicMisconduct}.
Academic misconduct refers to actions that violate the originality of academic work, such as ghostwriting, plagiarism, data fabrication, deceit, and generation using Artificial Intelligence (AI). Among these, generation using AI is the most frequent and recent form of misconduct \cite{nerdynavChatGPTCheating}. The research primarily focuses on this aspect. Recent generative AI tools such as ChatGPT, GPT-4 Vision, SORA from OpenAI, Gemini from Google, and Perplexity from Perplexity AI are capable of generating various types of content, including text, images, video, and code in multiple programming languages. The introduction of generative AI, particularly ChatGPT from OpenAI, has posed challenges in academia \cite{nerdynavChatGPTCheating}.

Students can utilize models like ChatGPT for various tasks, such as generating an essay for a given topic, brainstorming ideas, summarising textual contents, solving programming assignments, solving complex mathematical equations, etc. Furthermore, ChatGPT now offers a feature where users can upload a PDF file and the AI will read the document and generate relevant text based on its contents \cite{indianexpressChatGPTLets}. This has unfortunately made plagiarism even easier \cite{indiatodayChatGPTJust}. In addition to completing assignments, ChatGPT can assist with solving problems in various examinations, potentially raising concerns about the credibility of the examination process. Using different strategies, such as prompting, ChatGPT provides a range of suggestions for cheating in university exams \cite{preprintsExploringEthical}.  It has demonstrated the ability to pass rigorous examinations, including the US Medical Licensing Exam  \cite{abcBiggerThan}. ChatGPT has also shown promising results in other exams such as the bar examination, Scholastic Assessment Test (SAT), Graduate Record Examination (GRE), 2020 USA Biology Olympiad Semifinal Exam, college-level microbiology quiz, and Stanford Medical School clinical reasoning final \cite{businessinsiderGPT4Bar}. Moreover, tools like ChatGPT can handle complex programming assignments as well \cite{geeksforgeeksChatGPTComplete}.  

Students and researchers have been using ChatGPT to write research papers and get published \cite{washingtonpostProfessorsPublished}. Recent Google search results have proved the use of ChatGPT in scholarly publishing \cite{ChatGPT_google}, with publishers making policies to moderate the usage of ChatGPT in academic publishing \cite{theguardianScienceJournals}. Therefore, some major recent academic threats raised are using generative AI tools such as ChatGPT to submit school/college assignments, pass licensure examinations, and publish research papers.

\section{Existing Solutions to AI-Generated Plagiarism}\label{existingsolutions}
Academic cheating was a serious issue even before the rise of LLMs because of easy access to information and other works on the internet \cite{eysenbach2000report}. Different solutions for Author Identification \cite{kestemont2019overview}, and Plagiarism Detection \cite{patel2011evaluation} have been developed. 
The solutions developed are being used by many academic institutions and publishers \cite{turnitinTurnitinCelebrates}, but they are not completely robust \cite{d2023turnitin}. However, with the introduction of LLMs, the problem of academic misconduct is even more serious because of the different use cases of LLMs to perform academic misconduct, such as completing student assignments, passing online examinations, paraphrasing texts to escape plagiarism and generating academic content for research papers. Due to the different problems raised by LLMs, research studies have been performed on AIGC detection. This section discusses plagiarism and plagiarism detection, along with how LLMs have changed plagiarism detection in recent years. Similarly, the section discusses the datasets, algorithms and tools for AIGC detection and the evasion techniques developed to fool such AIGC detectors.
\subsection{Plagiarism Detection}

 The American Historical Association (AHA) formally defined plagiarism in 1987 as failure to acknowledge the work of another \cite{plag_book}. Plagiarism, a part of academic cheating, is much more widespread than usually recognised according to a few studies \cite{uowPlagiarismUniversity}. 
 

The typology of plagiarism may vary according to data type or level of obfuscation. Foltnek et al. \cite{foltynek2019academic} presented different typologies defined in several research papers and put forward a new typology for plagiarism according to the level of obfuscation as character-preserving plagiarism, syntax-preserving plagiarism, semantics-preserving plagiarism, idea-preserving plagiarism, and ghostwriting. 
Plagiarism may also be monolingual or cross-lingual; monolingual plagiarism is done in one language, and cross-lingual plagiarism is done in multiple languages \cite{ieeeUnderstandingPlagiarism}. The plagiarism type may also vary according to the data types, such as code plagiarism and text plagiarism. 

 Plagiarism detection also may be categorized based on different factors. Based on the number of languages used, plagiarism detection may be monolingual or cross-lingual. Likewise, plagiarism detection may be extrinsic or intrinsic. Suppose plagiarism is detected only using the text itself. In that case, it is termed intrinsic plagiarism detection, whereas if plagiarism is detected in comparison with other text, it is termed extrinsic plagiarism detection \cite{alsallal2019integrated}. 
Plagiarism detection can be categorised according to their approach. N-gram-based, vector-based, syntax-based, semantic-based, fuzzy-based, structural-based, and stylometric-based \cite{khaled2021plagiarism}.

As the internet grows larger day by day, with more textual information and new tools for plagiarism, LLMs are one of them, and the concern of plagiarism detection is even more serious. A large survey conducted for 12 years at 24 universities in the US presenting that almost 95\% of the students admit to plagiarism at least once \cite{argassociationPlagiarismStatistics} demonstrates the need for plagiarism detection. Many research conferences and workshops such as Plagiarism Analysis, Authorship Identification, and Near-duplicate Detection (PAN), and International Conference Plagiarism across Europe and Beyond \cite{chaika2023zero} have been held to solve plagiarism detection \cite{eriksson2012features, stein2011fourth, potthast2009overview}.
 Different datasets have been developed and made open source to solve plagiarism detection. The datasets, source type, publisher, and size highlighting the number of observations have been summarized in Table \ref{tab:plagiarism-datasets}.


\begin{table}[h]
    \centering
    \caption{Summary of different open source datasets developed to solve plagiarism detection along with the number of instances, source type, and publisher.}
    \label{tab:plagiarism-datasets}
    \rowcolors{2}{gray!15}{white}
    \begin{tabular}{p{6cm} p{3.5cm} p{2cm} p{3.5cm}}
        \rowcolor{gray!50}
        \toprule
        \textbf{Dataset} & \textbf{Number of Instances} & \textbf{Source Type} & \textbf{Publisher} \\
        \midrule
        Microsoft Research Paraphrase Corpus (MRPC) \cite{dolan2005automatically} & 5,801 & Paper & Microsoft \\
        Michigan Deep Blue Data \cite{foltynek2020detecting} & 4,012 & Paper & University of Michigan \\
        Cross-Lingual Plagiarism Detection Dataset \cite{stegmuller2021detecting} & 2,000 & Paper & Zenedo \\
        Machine Paraphrase Dataset \cite{huggingfacedataset} & 200,100 & Paper & HuggingFace \\
        Urdu Intrinsic Plagiarism Detection \cite{haseeb2024versatile} & 10,872 & Paper & Mendeley \\
        4lang \cite{avetisyan2023cross} & Multiple Datasets & Mendeley Data & Mendeley \\
        PAN Dataset \cite{eriksson2012features} & Multiple Datasets & PAN Series & Webis Group \\
        \bottomrule
    \end{tabular}
\end{table}


A typical plagiarism detection algorithm involves feature engineering, classification models or text-matching similarity metrics.
Frequency of characters, average word length, average sentence length, Word N-grams frequency, part of speech, synonyms, and hypernyms were common features used in most plagiarism detection algorithms \cite{chitra2016plagiarism}. Plagiarism is mainly evaluated based on textual similarity with other reference textual contents. To calculate such similarity, hamming distance, Levenshtein distance, and longest common subsequence distance were the most commonly used string similarity metrics, whereas Jaccard coefficient, Cosine coefficient, Manhattan distance, euclidean distance, Matching coefficient and Dice coefficient were the most commonly used vector similarity metrics \cite{alzahrani2011understanding}. 
The different algorithms developed along with the datasets, models, and evaluation metrics used for plagiarism detection are summarised in Table \ref{tab-plagiairism-detection-algorithms}.

\begin{table}[h]
    \centering
    \caption{Summary of different algorithms developed for plagiarism detection along with the datasets and models used and the metrics used for evaluation.}
    \label{tab-plagiairism-detection-algorithms}
    \rowcolors{2}{gray!15}{white}
    \begin{tabular}{p{5.5cm} p{1cm} p{3cm} p{3cm} p{2cm}}
        \rowcolor{gray!50}
        \toprule
        \textbf{Title} & \textbf{Year} & \textbf{Datasets Used} & \textbf{Models used} & \textbf{Metrics} \\
        \midrule
        A New Online Plagiarism Detection System based on Deep Learning \cite{el2020new} & 2020 & PAN dataset & Doc2Vec, SLSTM, CNN & Accuracy \\
        Plagiarism Detection Using Machine Learning-Based Paraphrase Recognizer \cite{chitra2016plagiarism} & 2015 & PMRPC dataset & Feature Extraction, SVM & Accuracy \\
        Will ChatGPT get you caught? Rethinking of Plagiarism Detection \cite{khalil2023will} & 2023 & Custom Data & ChatGPT & False Negatives \\
        Plagiarism Detection Using Machine Learning-Based Paraphrase Recognizer \cite{chitra2016plagiarism} & 2023 & Custom Dataset & GPT4 & False Negatives \\
        A Population-based Plagiarism Detection using DistilBERT-Generated Word Embedding \cite{yuqin2023population} & 2023 & SNLI, MRPC and SemEval 2014 & DistilBERT, LSTM, Clustering & Precision, recall, F1, G-means \\
        Arabic Plagiarism Detection Using Word Correlation in N-Grams with K-Overlapping Approach \cite{alzahrani2015arabic} & 2015 & Arabic Plagiarism Dataset & N Gram Similarity Matching & Recall, Precision, Granularity, Plagdet \\
        Exploration of Fuzzy C Means Clustering Algorithm in External Plagiarism Detection System \cite{riya2016exploration} & 2015 & PAN 2013 Corpus & Fuzzy C means Clustering Algorithm & Precision, Recall \\
        \bottomrule
    \end{tabular}
\end{table}


Along with such research studies, several plagiarism detection tools have been developed and made available online. Abdelhamid et al. \cite{abdelhamid2022survey} compared the performance of 8 different plagiarism detectors online on different levels of plagiarism in text. Among plagiarism detectors online, Turnitin is the tool mostly favoured by academic institutions \cite{mphahlele2019use}. 
The most commonly used tools, along with their important feature, type of licensing, and release years, have been summarised in Table \ref{tab-plag-detection-tools}.

    

\begin{table}[h]
    \centering
    \caption{Summary of the most popular tools available online for detecting plagiarism.}
    \label{tab-plag-detection-tools}
    \rowcolors{2}{gray!15}{white}
    \begin{tabular}{p{3cm} p{2.5cm} p{1.5cm} p{8cm}}
        \rowcolor{gray!50}
        \toprule
        \textbf{Tool Name} & \textbf{Release Year} & \textbf{Licensing} & \textbf{Feature} \\
        \midrule
        Turnitin & 2000 & Paid & Supports intra-corporal and extra-corporal detection\\
        iThenticate & 1996 & Paid & Support multiformat document \\
        CopyLeaks & 2015 & Paid & Support multiple file formats  \\
        Duplichecker & 2020 & Free & Can copy-paste or upload documents to check plagiarism \\
        Grammarly & 2016 & Paid & Shows the plagiarism source \\
        PlagScan & 2009 & Free & Handle multiple types of files, compare side by side review\\
        \bottomrule
    \end{tabular}
\end{table}

The major problem with protecting originality in academic settings is the new evasion techniques that people come up with to avoid existing plagiarism detection tools. Elkhatat et al. \cite{elkhatat2021some} experimented with four of the most commonly used evasion techniques applied to top plagiarism detectors available online and found plagiarism detectors cannot detect simple evasion techniques. The evasion techniques that students mostly use to fool existing online plagiarism detection tools, along with their short description and the result of checking those tools through every technique, are presented in Table \ref{plagiarism_evasion_techniques}. 


\begin{table}[h]
    \centering
    \caption{Summary of Evasion Techniques for Plagiarism Detection and Result while Testing on Plagiarism Detectors.}
    \label{plagiarism_evasion_techniques}
    \rowcolors{2}{gray!15}{white}
    \begin{tabular}{p{3cm} p{4cm} p{4cm} p{4cm}}
        \rowcolor{gray!50}
        \toprule
        \textbf{Evasion Technique} & \textbf{Description} & \textbf{Successful Plag Detectors} & \textbf{Unsuccessful Plag Detectors} \\
        \midrule
        Imaged Texts & Convert all text to image and export files as PDF & None & Turnitin, iThenticate, Copyscape, PlagAware, StrikePlagiarism.com, Unicheck, Check-For-Plag, Blackboard-SafeAssign \\
        Quoted & Insert invisible white quotation marks for all paragraphs & Blackboard-SafeAssign, Copyscape, PlagAware, StrikePlagiarism.com, Unicheck, Check-For-Plag & Turnitin, iThenticate, PlagScan \\
        Letter-like Symbols & Replace most common letters like a, e, o with their Latin small letter & Turnitin, iThenticate, PlagAware & Blackboard-SafeAssign, Copyscape, PlagScan, Unicheck, Check-For-Plag \\
        Invisible Letters & Replace spaces within words with any white color letter & StrikePlagiarism.com & Turnitin, iThenticate, Copyscape, PlagAware, Unicheck, Check-For-Plag, Blackboard-SafeAssign \\
        \bottomrule
    \end{tabular}
\end{table}

\subsubsection{Plagiarism Detection After LLMs}

Plagiarism and its Detection have changed their course since the rise of LLMs. 
Although many paraphrasing tools exist, such as Wordtune and Quillbot, students can also use LLMs to paraphrase the given text and attempt to escape plagiarism. On the other hand, LLMs can also be used as a plagiarism detection tool. There are several commercial tools to check plagiarism scores. However, students can simply use ChatGPT for free to check whether a text is plagiarised. This helps the students estimate the likelihood of getting caught and manipulate the copied text to fool plagiarism detection tools. Birock et al. \cite{divaportal} performed four prompt engineering experiments to determine the efficiency of ChatGPT as a plagiarism testing tool and found the fourth prompt to work very well. In another experiment by Khali et al. \cite{khalil2023will} on 50 essays generated by ChatGPT and tested, ChatGPT showed superior performance in plagiarism detection than other traditional plagiarism detection tools like iThenticate.

\begin{figure}[hbt!]
    \centering
    \includegraphics[width=150mm]{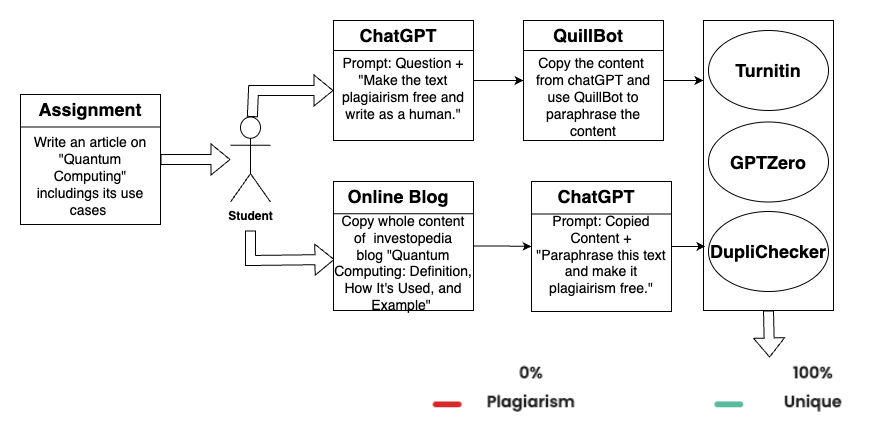}
    \caption{Diagram demonstrating how ChatGPT and paraphrasing tools can be used to complete assignments.}
    \label{fig:ChatGPTparaphrasing}
\end{figure}

Students can also use both ChatGPT and paraphraser to fool the existing plagiarism detection and AIGC detection tools.  Instead of copying the content written by others and paraphrasing the content using several tools to fool plagiarism detectors, students can simply generate textual content from ChatGPT and apply paraphrasing to make the content non-AIGC detectable and plagiarism-free. To demonstrate this, we conducted a simple experiment in which a student was assigned to write an article on quantum computing, including its use cases. We generated the article using ChatGPT, paraphrased the generated text using QuillBot, and tested its originality. Similarly, we copied an article from an online blog from Investopedia and copied that content to ChatGPT to paraphrase and again tested the originality of the text. The testing used the plagiarism detection tool Turnitin and the AIGC detection tools  GPTZero and DupliChecker. The output text was identified as being 100\% unique, demonstrating that students can easily complete their assignment using such a combination of tools, and lecturers will not be able to find out. This experimentation is demonstrated in Figure \ref{fig:ChatGPTparaphrasing}.

\subsection{AI Generated Content (AIGC) Detection}

The major objective of AIGC Detection is detecting whether a given piece of text is generated using an AI system or written by a human. It is very difficult even for a human to differentiate because of the high capability of LLMs nowadays to produce more and more human-like text. In an experiment, English instructors hired to differentiate AI-generated and human-written essays were only able to get 67\% accuracy \cite{liu2023argugpt}. It is more difficult to distinguish text generated by language models from human-written text compared to distinguishing text generated by different language models \cite{shijaku2023ChatGPT}. In an experiment, AI models were able to generate more high-quality argumentative essays than German high school students in an online forum \cite{herbold2023large}. Similarly, in another experiment, six undergraduate and PhD students were asked to annotate 50 documents as AI-generated or human-written. They could only achieve 59\%  accuracy \cite{verma2023ghostbuster}. Thus, AIGC detection is a very complex task. 

\begin{figure}[hbt!]
    \centering
    \includegraphics[width=120mm]{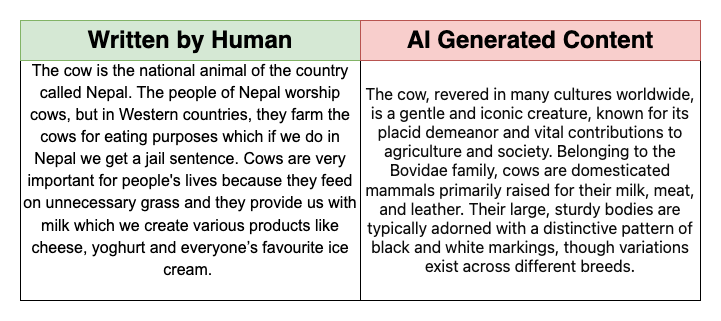}
    \caption{Example of a ChatGPT generated and human-written text.}
    \label{fig:ChatGPTvshuman}
\end{figure}

Human essays have more spelling and grammar errors and personal experiences. In contrast, machine essays have more similar examples and repetitive expressions, according to a quantitative analysis performed between human essays and ChatGPT-written essays \cite{liu2023argugpt}. Machine-generated text has a more complex syntactic structure and uses more normalization, whereas human essays tend to be more lexically complex \cite{liu2023argugpt,herbold2023large}. Similarly, human-written text tends to have higher perplexity compared to AI-generated text \cite{liao2023differentiate}. To demonstrate such linguistic differences in human-written and AI-generated text, a human was told to write a paragraph on "cow", and the prompt "Write a paragraph on Cow" was injected in ChatGPT. The grammatical mistakes and use of personal experience can be seen in the human-written text. In contrast, syntactically complex sentence structure and repetition of words can be seen in the text generated from ChatGPT, shown in Figure \ref{fig:ChatGPTvshuman}.

\subsubsection{Open Source Datasets for AIGC Detection}
As the term "Garbage In, Garbage Out" in machine learning represents, high-quality data is vital for efficient machine learning research \cite{geiger2020garbage}. AIGC detection, when modelled as a straightforward binary classification problem, requires a high-quality, balanced labelled dataset comprising a wide variety of human-written and AI-generated texts. However, when AIGC detection is solved from other approaches, such as zero-shot detection methods, a dataset may not be required for training \cite{xian2017zero}. Building a high-quality dataset is a very laborious process that involves data scraping from various sources, data cleaning, data preprocessing, and data annotation. The human-written part of most open-source AIGC detection datasets was built using three major techniques. First is directly taking from other open-source datasets such as the wikiHow text dataset \cite{koupaee2018wikihow}. This technique was used to build the GPT-Sentinel dataset \cite{chen2023gpt}. Second is using an available corpus such as in-class or homework exercises, Test of English as a Foreign Language (TOEFL) writing tasks, and GRE writing tasks were used to build ArguGPT dataset \cite{liu2023argugpt}. Third is manually collecting observations from online sources such as the CHEAT dataset built by searching and extracting human-written abstracts from IEEE Xplore \cite{yu2023cheat}.
On the other hand, the AI-generated text part in those open-source datasets is built using different prompting strategies applied to OpenAI models using the OpenAI API service \cite{chen2023gpt,yu2023cheat}. Some papers claim very high results in AIGC detection using specific custom datasets, but the dataset is not open-sourced \cite{chen2023gpt}. The datasets used in different research papers to solve AIGC detection are presented in Table \ref{datasets_AIGC}.

     

\begin{table}[h]
    \centering
    \caption{Summary of different open-source datasets developed for AIGC detection along with the source of the human-written data part, number of instances, and publisher.}
    \label{datasets_AIGC}
    \rowcolors{2}{gray!15}{white}
    \begin{tabular}{p{3.5cm} p{2.5cm} p{6cm} p{3cm}}
        \rowcolor{gray!50}
        \toprule
        \textbf{Dataset} & \textbf{Number of Instances} & \textbf{Human Written Text Source} & \textbf{Publisher} \\
        \midrule
         ArguGPT \cite{arxivArguGPTEvaluating} & 4,708 & Student essays from TOEFEL  & \href{https://huggingface.co/datasets/SJTU-CL/ArguGPT}{HuggingFace} \\
        Custom Dataset & 252 & Student essays from TOEFEL & \href{https://github.com/rexshijaku/ChatGPT-generated-text-detection-corpus}{Github}\\
        Human ChatGPT comparison corpus (HC3) \cite{guo-etal-2023-hc3} & 24,300 & Open-domain, financial, medical, legal, and psychological areas & \href{https://huggingface.co/datasets/Hello-SimpleAI/HC3}{HuggingFace} \\
        CHEAT \cite{yu2023cheat} & 35,304 & Abstracts from papers & \href{https://paperswithcode.com/dataset/cheat}{PaperswithCode}\\
        DagPap22 \cite{fake-papers-competition-data-repo} & 26,637 & Elsevier papers & \href{https://github.com/Yorko/fake-papers-competition-data}{Github}\\
        Ghostbuster Dataset \cite{arxivGhostbusterDetecting} & 12,500 & Subreddit posts & \href{https://github.com/vivek3141/ghostbuster-data/tree/master}{Github} \\
        M4GT-Bench Dataset \cite{wang2024m4gt} & 138,465 & Wikipedia, Reddit, Arxiv, and PeerRead & \href{https://arxiv.org/pdf/2402.11175.pdf}{Arxiv} \\
        M4 Dataset \cite{wang-etal-2024-m4} & 147,895 & Wikipedia, Reddit, Arxiv, PeerRead, Urdu-news, and RuATD & \href{https://github.com/mbzuai-nlp/M4}{Github} \\
        GPT-Sentinel \cite{arxivGPTSentinelDistinguishing} & 3,152,979 & OpenGPTText, OpenWebText and ChatGPT & \href{https://arxiv.org/pdf/2305.07969.pdf}{Arxiv} \\
        \bottomrule
    \end{tabular}
\end{table}

A few gaps were found in this survey of the datasets used for AIGC detection. Firstly, no benchmark dataset is specified for the problem. Researchers have been developing custom data on their own \cite{arxivArguGPTEvaluating, chen2023gpt}. The evaluation/testing dataset is also made from their dataset, making the comparison between these research implementations impartial. Secondly, the different datasets built by different researchers contain textual data from a specific domain.
Secondly, the datasets built contain straightforward ChatGPT-generated content. Due to this, several techniques like paraphrasing can fool the AI detectors \cite{sadasivan2023can}. To address these gaps, a benchmark dataset with the proper proportion of text from a wide range of fields and text observations which contain modified ChatGPT answers should be developed, and solutions developed for AIGC detection should be evaluated based on the benchmark data evaluation. 

\subsubsection{Watermarking Based Approaches }
Watermarking-based approaches embed subtle signatures like a cryptographic pseudo-random function in LLM-generated text, which can be decrypted further to check whether the text is generated by the particular LLM. The input and output in text generation models like LLMs are always tokens. LLMs constantly generate a probability distribution over the next predicted token, conditional on the previous string of tokens. Based on such probability distribution, the next token is sampled randomly according to a parameter called temperature. Now, instead of randomly selecting the next token in the output, a cryptographic pseudorandom function can be introduced, which is unnoticeable to the end users but can be further decrypted to see whether that particular LLM generates a given text or not. 

Watermarking in natural language was done for other purposes like information hiding even before the release of ChatGPT \cite{topkara2006hiding}. Techniques like morphosyntactic alterations and synonym substitution were used for watermarking natural language \cite{meral2009natural, hao2018reversible}. In a workshop on LLMs and Transformers, Scott Aaronson revealed that he proposed a watermarking scheme based on the "Gumbel Softmax Rule", whose prototype was later implemented by another scientist from OpenAI, Hendrick Kirchner which seems to work even with a few hundred tokens \cite{scottaarosnsonblog}. The problem with these traditional watermarking techniques was that they could not preserve the semantic meaning between previous and watermarked texts, i.e., the original and watermarked texts had very different meanings. Yang et al. \cite{yang2022tracing} introduced context-aware lexical substitution for watermarking and achieved a 2.19 Z-score which represents the watermark strength. Higher z-scores indicate that z-scores of these features in the watermarked text deviate from the mean values found in human-written text in a controlled manner.  Sahar Abdelnabi and Mario Fritz  \cite{abdelnabi2021adversarial} introduced the first end-to-end Adversarial Watermarking Transformer (AWT) model just a month after the GPT-2 release. AWT model was able to achieve a z-score of 2.73. Yang et al. \cite{yang2023watermarking} again developed another framework for watermarking black box language models with a z-score of 3.63. 

Kirchenbauer et al. \cite{kirchenbauer2023watermark} proposed a watermarking technique in June 2023. This algorithm reduced the False Positive Rate (FPR) to 0 and could be used without prior knowledge of any LLM parameters or its Application Programmable Interface (API). However, in October 2023, Zhao et al. \cite{zhao2023provable} proved that the Kirchenbauer et al.  \cite{kirchenbauer2023watermark} watermarking technique failed when a paraphrasing attack was performed using ChatGPT, DIPPER-1, DIPPER-2 and BART. Along with these results, Zhao et al. \cite{zhao2023provable} also proposed another watermarking solution called Unigram-Watermark, which outperformed the previous solution proposed by Kirchenbauer et al. \cite{kirchenbauer2023watermark} for paraphrasing attacks. Still, the watermarking techniques faced the challenge of maintaining the detectability of inserted watermarks and semantic integrity of generated text \cite{huo2024token}. To solve this, Huo et al. \cite{huo2024token} proposed a multiobjective optimization (MOO) approach for watermarking, which outperformed previous watermarking techniques and was robust against copy-paste attacks and paraphrasing attacks. Again, in March 2024, a new watermarking framework, WaterMax, was introduced by Giboulot et al.  \cite{giboulot2024watermax}, which results in high detectability, outperforming previous watermarking techniques and is also designed in such a way that it leaves the LLM untouched, maintaining the quality of the generated text. A summary of different research papers for AIGC detection based on the watermarking technique is represented in Table \ref{tab_watermarking}.


\begin{table}[h]
    \centering
    \caption{Summary of AIGC Detection Research Papers Based on Watermarking along with the metrics used to evaluate the approach.}
    \label{tab_watermarking}
    \rowcolors{2}{gray!15}{white}
    \begin{tabular}{p{7cm} p{2.5cm} p{1cm} p{4cm}}
        \rowcolor{gray!50}
        \toprule
        \textbf{Paper} & \textbf{Authors} & \textbf{Year} & \textbf{Evaluation Metrics} \\
        \midrule
        Watermarking Text Generated by Black-Box Language Models \cite{yang2023watermarking} & Xi Yang et al. & 2023 & ROC, AUC, Semantic similarity, METEOR score \\
        Tracing Text Provenance via Context-Aware Lexical Substitution \cite{yang2022tracing} & Xi Yang et al. & 2021 & Semantic Relatedness (SR), Semantic Similarity (SS) \\
        Adversarial Watermarking Transformer: Towards Tracing Text Provenance with Data Hiding \cite{abdelnabi2021adversarial} & Sahar Abdelnabi and Mario Fritz & 2022 & Bit accuracy, METEOR score, SBERT distance \\
        A Watermark for Large Language Models \cite{kirchenbauer2023watermark} & John Kirchenbauer et al. & 2023 & FPR, TNR, TPR, FNR \\
        Provable Robust Watermarking for AI-Generated Text  \cite{zhao2023provable} & Xuandong Zhao et al. & 2023 & TPR, F Score \\
        Token-Specific Watermarking with Enhanced Detectability and Semantic Coherence for Large Language Models  \cite{huo2024token} & Mingjia Huo et al. & 2024 & FPR, Semantic coherence \\
        WaterMax: breaking the LLM watermark detectability-robustness-quality trade-off  \cite{giboulot2024watermax} & Eva Giboulot, Teddy Furon & 2024 & P values, ROC, tamper resistance \\
        \bottomrule
    \end{tabular}
\end{table}

\subsubsection{Zero-shot Based Approaches}
Zero-short approaches are those in which pretrained neural network models can predict unseen classes \cite{joeddavZeroShotLearning}. While typical text classification models predict whether an observation falls into a specific class or not after training the model on a huge set of labelled datasets, the zero-shot learning approach utilizes the ability of pretrained language models to generalize unseen text observations or even new datasets \cite{pushp2017train}. 

Giant Language Model Test Room (GLTR) is a pioneering work in AIGC detection based on zero-shot learning. Gehrmann et al. \cite{gehrmann2019gltr} developed GLTR that can detect whether a text is generated by a model or not with a proper visual footprint of generated text through the tool that also proves the prediction. Mireshghallah et al. \cite{mireshghallah2023smaller} experimented to test whether other language models can be used to detect machine-generated text from one language model and found out that smaller language models like OPT-125M are better AIGC detectors than large language models like GPTJ-6B.
DNA-GPT was introduced in the AIGC detection space, which claimed to surpass the result of OpenAI text detector \cite{openaiClassifierIndicating} on four English and German datasets \cite{yang2023dna}. The main idea of DNA-GPT is to compare the probability divergence between the actual tokens and generated tokens \cite{yang2023dna}. 

Detect-GPT was another milestone in AIGC detection because it raised the state-of-the-art zero-shot detection of AIGC detection from the previous highest  0.81 AUROC to 0.95 AUROC \cite{mitchell2023detectgpt}. Detect-GPT was built on the principle that machine-generated text mostly occupies negative curvature regions of log probability generated from the model \cite{mitchell2023detectgpt}. Recently, Fast-DetectGPT increased the AUROC to 0.98 and obtained a speed 340 times faster compared to DetectGPT \cite{bao2023fast}. Fast-DetectGPT introduced the notion of conditional probability curvature to clarify disparities in vocabulary selections observed between machine-generated text and human-written text \cite{bao2023fast}. A summary of the research papers on AIGC detection based on zero-shot learning, along with their authors, published year and evaluation metrics, is represented in Table \ref{zero_shot}.


\begin{table}[h]
    \centering
    \caption{Summary of AIGC Detection Research Papers Based on Zero-Shot Learning along with the metrics used for evaluation by the approach.}
    \label{zero_shot}
    \rowcolors{2}{gray!15}{white}
    \begin{tabular}{p{8cm} p{3.5cm} p{1cm} p{2cm}}
        \rowcolor{gray!50}
        \toprule
        \textbf{Paper} & \textbf{Author} & \textbf{Year} & \textbf{Evaluation Metrics} \\
        \midrule
        GLTR: Statistical Detection and Visualization of Generated Text  \cite{gehrmann2019gltr} & Sebastian Gehrmann et al. & 2019 & AUROC \\
        Smaller Language Models are Better Black-box Machine-Generated Text Detectors \cite{mireshghallah2023smaller} & Niloofar Mireshghallah et al. & 2023 & AUC \\
        DNA-GPT: Divergent N-Gram Analysis for Training-free Detection Of GPT-Generated Text \cite{yang2023dna} & Xianjun Yang & 2023 & AUROC, TPR \\
        DetectGPT: Zero-Shot Machine-Generated Text Detection using Probability Curvature  \cite{mitchell2023detectgpt} & Eric Mitchel et al. & 2023 & AUROC \\
        Fast-DetectGPT: Efficient Zero-shot Detection of Machine-Generated Text via Conditional Probability Curvature  \cite{bao2023fast} & Guangsheng Bao et al. & 2024 & AUROC, Speed \\
        Does DETECTGPT Fully Utilize Perturbation? Bridge Selective Perturbation to Fine-tuned Contrastive Learning Detector would be Better \cite{mao2024raidar} & Shengchao Liu et al. & 2024 & Accuracy, F1 Score \\
        Spotting LLMs with Binoculars" Zero-shot detection of machine-generated text \cite{hans2024spotting} & Abhimanyu Hans et al. & 2024 & Precision, F1 Score, FPR \\  
        \bottomrule
    \end{tabular}
\end{table}

\subsubsection{Training Classifier Based Approaches}
AIGC detection can be framed as a binary text classification problem where we have the given text as input and AI-generated and human-written as two classes. After the release of ChatGPT in November 2022, OpenAI itself came up with an OpenAI Text Classifier in Jan 2022 \cite{techcrunchOpenAIReleases}.
Liu et al. \cite{liu2023argugpt} introduced ArguGPT a classification model trained on sentence-level and essay-level data from TOEFL and GRE tasks achieving 90\% accuracy. Oghaz et al. \cite{oghaz2023detection} implemented different machine-learning algorithms like Multinomial Naive Bayes, Random Forest, Support Vector Machines (SVM), and K-nearest neighbours (KNN) as well as deep learning algorithms such as Bidirectional Long Short Term Memory (LSTM) networks, DistilBERT, and  RoBERTa on a custom dataset of question answers mainly on computer science, artificial intelligence, and cyber security. In his experiment, the Roberta-based custom deep learning model achieved the highest performance with an F-score of 0.992 and an accuracy of 0.991 \cite{oghaz2023detection}.

Most text classification algorithms have applied common text preprocessing techniques such as tokenization, stemming, lowercasing, and stopword removal. Similarly, TFIDF has been used for feature extraction from the text \cite{shijaku2023ChatGPT}. Hijaku et al. \cite{shijaku2023ChatGPT}implemented the XGBoost algorithm with TFIDF and handcrafted features, which were further visualized using SHAP analysis. Katib et al. \cite{katib2023differentiating} introduced a new approach to train a binary classification model for AIGC detection, the Tunicate Swarm Algorithm with Long Short-Term Memory Recurrent Neural Network (TSA-LSTMRNN). The new approach achieved 93.17\% F-score and 93.83\% accuracy on human- and ChatGPT-generated datasets, respectively \cite{katib2023differentiating}. This experimentation was done on the CHEAT dataset. Wissam Antoun et al. \cite{antoun2023towards} implemented different transformer-based algorithms such as Roberta, ELECTRA, CamemBERTa, and XLM-R on a modified Human ChatGPT Comparision Corpus (HC3) dataset containing English and French text on which a 99.86 F-Score was achieved. GPT-Sentinel was another pioneer work based on binary classification using transformer-based networks like the Robustly Optimized BERT Pretraining Approach (Roberta) and Text-to-Text Transfer Transformer (T5) achieving over 97\% accuracy \cite{chen2023gpt}.  The current state-of-the-art classification algorithm is the Ghostbuster algorithm, which claims to achieve  99.0 F1, which is 5.9 F1 higher than the best pre-existing model \cite{verma2023ghostbuster}. A summary of research papers for AIGC detection based on training a classifier is represented in Table \ref{training_based}.

\begin{table}[h]
    \centering
    \caption{Summary of AIGC Detection Research Papers Based on Training Classifier along with the datasets and models used for training.}
    \label{training_based}
    \rowcolors{2}{gray!15}{white}
    \begin{tabular}{p{6cm} p{2cm} p{1cm} p{2.5cm} p{3cm}}
        \rowcolor{gray!50}
        \toprule
        \textbf{Title} & \textbf{Author} & \textbf{Year} & \textbf{Dataset Used} & \textbf{Models Used}\\
        \midrule
        ArguGPT: evaluating, understanding and identifying argumentative essays generated by GPT models \cite{liu2023argugpt} & Yikang Liu et al. & 2023 & Custom Essay Dataset & SVM, Roberta \\
        Comparing scientific abstracts generated by ChatGPT to original abstracts using an artificial intelligence output detector, plagiarism detector, and blinded human reviewers  \cite{gao2022comparing}& Catherine A. Gao et al. & 2023 & Custom Dataset using abstracts of research papers & ChatGPT \\
        Detection and Classification of ChatGPT Generated Contents Using Deep Transformer Models \cite{oghaz2023detection} & Mahdi Maktab et al.R & 2023 & custom dataset & Machine Learning Algorithms, BiLSTM, DistilBERT, Roberta \\
        Ghostbuster: Detecting text written by large language modes  \cite{verma2023ghostbuster} & Vivek Verma & 2023 & Custom Dataset & LLMs \\
        LLM-Detector: Improving AI-Generated Chinese Text Detection with Open-Source LLM Instruction Tuning \cite{wang2024llm} & Rongsheng Wang et al. & 2024 & Custom Chinese Dataset & GPT \\
        Paraphrasing evades detectors of AI-generated text, but retrieval is an effective defence \cite{arxivParaphrasingEvades} & Kalpesh Krishna et al. & 2023 & Custom Dataset & DIPPER, open source LLMs \\
        Towards a Robust Detection of Language Model-Generated Text: Is ChatGPT that Easy to Detect? \cite{antoun2023towards} & Wissam Antoun et al. & 2023 & HC3 Corpus & RoBERTa, ELECTRA, CamemBERT, CamemBERTa\\
        \bottomrule
    \end{tabular}
\end{table}

\subsubsection{Detection Tools}

With the development of various algorithms to solve the AIGC detection problem, several tools have been made public utilizing those state-of-the-art models and algorithms. OpenAI itself announced the release of the OpenAI text detector, which was shut down later because of its low accuracy rate \cite{openaiClassifierIndicating}. Dreamsoft Innovations also introduced GPTKit in the market, which claimed to have 93\% accuracy \cite{GPTKit}. Later, DetectGPT was also released along with the Application Programmable Interface (API) for developers. Originality.ai claims to be the most accurate AI detector for different LLMs like ChatGPT, GPT-4, Bard, Claude 2 and others \cite{OriginalityAI}. Arslan Akram \cite{akram2023empirical} tested six different AIGC detection tools with 11,580 testing samples from different datasets and concluded that the originality.ai tool was particularly effective across the test dataset. Some of the most commonly used tools to detect text generated from LLMs, along with their accuracy obtained from the experiment conducted by Akram \cite{akram2023empirical}, are presented in Table 
 \ref{AIGC_tools}.

While most of the research papers for AIGC detection focus on simply giving output as AI-generated or not, some available online AIGC detectors also give results in the probability of a particular text being human-written or generated from AI. We tested the results from the AI detectors from Copyleaks, ZeroGPT, Quillbot, Writer, GPTZero, Detecting-AI, Sapling, Undetectable, and Crossplag. Excluding the AI detector of Copyleaks, all the other AI detectors gave the output in probability. The AI Detector of Quillbot also classified the given text in terms of AI-generated, AI-generated + paraphrased, human-written and human-written + paraphrased. Undetectable also had the feature to humanize any AI-generated text. The Detecting-AI outputs the prediction from different prediction models. 


\begin{table}[h]
    \centering
    \caption{Summary of tools available online for AIGC detection along with their availability, key features, accuracy reported and API availability.}
    \label{AIGC_tools}
    \rowcolors{2}{gray!15}{white}
    \begin{tabular}{p{2cm} p{3cm} p{6cm} p{1.5cm} p{2cm}}
        \rowcolor{gray!50}
        \toprule
        \textbf{Tool} & \textbf{Availability} & \textbf{Feature} & \textbf{Acc (\%)} & \textbf{API Availability} \\
        \midrule
        GPTkit  & Free up to 2048 characters & Utilizes 6 different AI-based content detection techniques for high accuracy & 88 & Yes \\
        GPTZero  & Free for 10,000 words per month & Can detect text from ChatGPT, GPT4, Bard, Llama and other AI models & 40 & Yes \\
        Originality \cite{https://originality.ai/} & \$0.01/100 words & Can submit longer text, detailed highlighting, faster scans & 96 & Yes \\
        Sapling & 1 million chars at \$25/month & Shows percentage of AI generation, also has extension & 60 & Yes \\
        Writer & Free  & Provides Software Development Kit (SDK) for developers & 48 & No \\
        Zylalab  & \$24.99 per month & Makes use of OpenAI technology & 55 & Yes \\
        \bottomrule
    \end{tabular}
\end{table}

\subsubsection{Techniques to evade AIGC Detection Tools}
With the development of different AIGC detection algorithms, people have also come up with different evasion techniques and algorithms to fool such AIGC detectors. These evasion techniques vary with the solution applied for AIGC detection. For instance, for AIGC detection systems against watermarking, techniques such as watermark stealing \cite{jovanovic2024watermark}, Self Color Testing-based Substitution (SCTBS) \cite{wu2024bypassing}, and language translation \cite{he2024can} have been developed. Similarly, paraphrasing, word substitution, and sentence substitution are used to evade binary classification training-based AIGC detection models \cite{peng2024hidding} \cite{krishna2024paraphrasing}. Other most common evasion techniques used to fool AIGC detectors are prompting \cite{lu2023large}, adversarial attacks \cite{peng2024hidding}, single space adding technique \cite{cai2023evade} and reinforcement learning \cite{nicks2023language}. These evasion techniques are tested on different AIGC detection models and have successfully fooled those models, resulting in a high drop in AUC and detection accuracy, presented in Table \ref{AIGC_evasion}.

\begin{table}[h]
    \centering
    \caption{Summary of different research papers on AIGC evasion techniques along with evasion technique applied, tested models, and results of the evasion applied.}
    \label{AIGC_evasion}
    \rowcolors{2}{gray!15}{white}
    \begin{tabular}{p{5cm} p{3cm} p{3.5cm} p{3.5cm}}
        \rowcolor{gray!50}
        \toprule
        \textbf{Experiment} & \textbf{Evasion Technique Applied} & \textbf{Tested Models} & \textbf{Results} \\
        \midrule
        Hiding the Ghostwriters: An Adversarial Evaluation of AI-Generated Student Essay Detection \cite{peng2024hidding} & Paraphrasing, Word Substitution, Sentence Substitution & ArguGPT, CheckGPT, RoBERTA-Single, RoBERTa-QA, Quality & Significant drop in accuracy and AUC on all models \\
        Bypassing LLM Watermarks with Color-Aware Substitutions \cite{wu2024bypassing} & Self Color Testing-based Substitution & Watermarked LLMs & SCTS evasion technique is highly effective for different watermarking schemes \\
        Watermark Stealing in Large Language Models \cite{jovanovic2024watermark} & Watermark Stealing & Watermarked LLMs & For under \$50 an attacker can both spoof and scrub state-of-the-art watermarking schemes with a success rate over 80\% \\
        Paraphrasing evades detectors of AI-generated text, but retrieval is an effective defence \cite{arxivParaphrasingEvades} & Paraphrasing, Word Substitution, Sentence Substitution & Watermarking, GPTZero, DetectGPT and OpenAI text Classifier & Paraphrasing dropped detection accuracy of DetectGPT from 70.3\% to 4.6\% \\
        Large Language Models can be Guided to Evade AI-Generated Text Detection \cite{lu2023large} & Substitution-based In-Context example Optimization method (SICO) & GPT3-D, DPT2-D, GPTZero, OpenAI-D, DetectGPT, Log\_Rank & Decreased the AUC of six detectors by 0.5 on average \\
        Language Model Detectors are easily optimized against \cite{nicks2023language} & Reinforcement Learning & OpenAI RoBERTa-Large detector & Decreased the AUC of OpenAI detector from 0.84 to 0.62 \\
        Stumbling Blocks: Stress Testing the Robustness of Machine-Generated Text Detectors Under Attacks \cite{wang2024stumbling} & Editing, Paraphrasing, Prompting, and Co-generating & Watermark, OpenAI Detectors, GLTR, DetectGPT & Performance drops by 35\% across all attacks \\
        Evade ChatGPT detectors via a single space \cite{cai2023evade} & Add a single-space Space Strategy & ChatGPT, GPTZero, GPT-4 & increased the evasion rate close to 100\% \\
        \bottomrule
    \end{tabular}
\end{table}

\begin{figure}[hbt!]
    \centering
    \includegraphics[width=15cm, height=150mm]{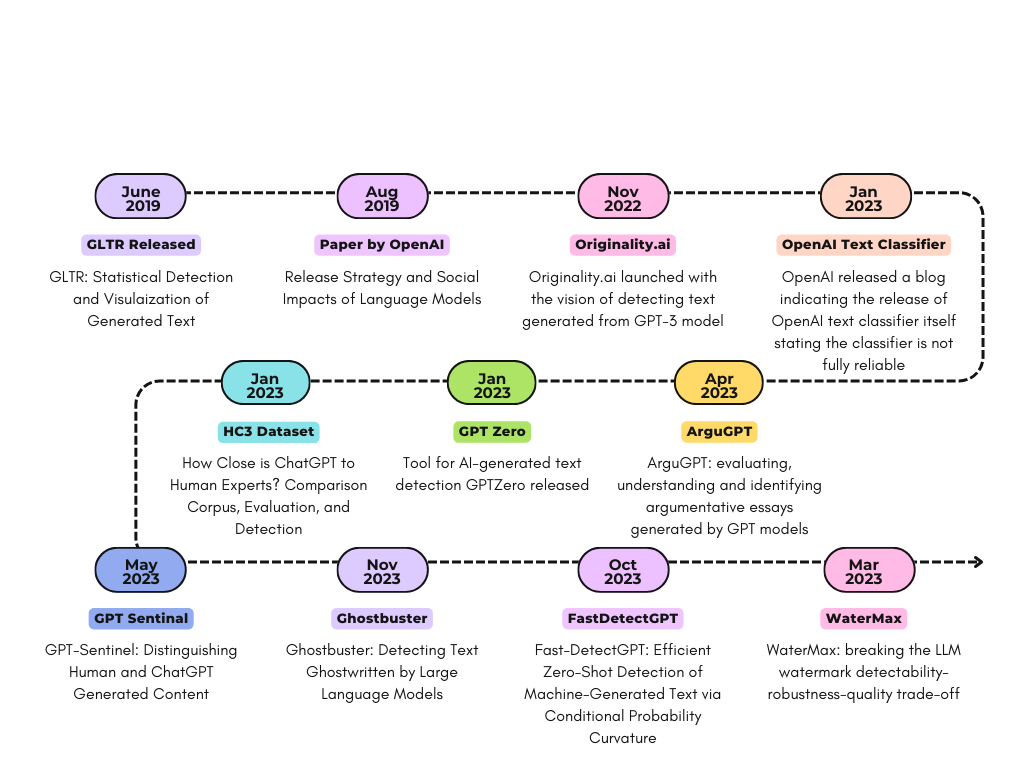}
    \caption{Major AIGC Detection Events including the description of top AIGC detection datasets, algorithms, and tools.}
    \label{fig:AIGC-events-label}
\end{figure}

The major events in the AI-generated text detection space are highlighted by the timeline represented in Figure \ref{fig:AIGC-events-label}. 

\section{Limitations and Gaps in Current Solutions}\label{gaps}
Although several solutions for AIGC detection have been developed, they can not be relied upon. Several cases of false positives have been reported \cite{turnitinUnderstandingFalse}, which questions the feasibility of developing a completely reliable solution. From this extensive survey, several gaps were identified in the existing solutions, which, if further research, can lead to a reliable solution for the problem.

\subsection{Current Scenario and Reliability of AIGC Detection}
As reported by Turnitin in 2023, over 16,000 academic institutions, publishers, and corporations use the Turnitin software to prevent plagiarism \cite{turnitinTurnitinCelebrates}. However, 97\% of the institutions have not implemented any formal policy on using AI tools by students, and 71\% of the instructors have never used AI writing tools  \cite{tytonpartners}. On the other hand, 51\% of the students responded to continue using generative AI tools even if prohibited by the institution \cite{tytonpartners}. A survey among 1147 instructors reported that most of the instructors permitted the use of generative AI writing tools for brainstorming ideas for an assignment, helping edit writing and outlining a structure for an assignment. In contrast, the instructors were against using generative AI writing tools for writing a small and large part of an assignment or even an entire assignment \cite{tytonpartners}. The data indicates that every organisation should develop and regulate a formal policy on using generative AI writing tools. Along with such a policy, an efficient and reliable tool that detects AI-generated content is essential. Turnitin launched the AI writing detection tool on April 4, 2023, and since over 50 million submissions have been processed globally \cite{turnitinTurnitinCelebrates}. However, Turnitin has acknowledged the misclassification of human-written text as AI-generated or false positives. Such false positives have serious implications for using such tools in real settings. 


\subsection{Gaps Identified in Existing Solutions}
There were several gaps identified in existing solutions for academic misconduct. From the perspective of datasets built and used for AIGC detection, most algorithms that come up for  AIGC detection are built using text observations that are copy-pasted from LLM tools like ChatGPT. Most of the implementations are tested on naive datasets, i.e. datasets which contain human-written and direct ChatGPT-generated text, resulting in very high evaluation metrics but poor performance in real settings. The existing solutions claim high accuracy on the dataset because most of these datasets are built with human-written and text generated from LLMs directly. However, in real settings, students apply several evasion techniques like paraphrasing to fool the detectors. This is also because the models developed have not been cross-tested against other datasets. Thus, a benchmark dataset for AIGC detection representing those observations in which evasion techniques like paraphrasing have been performed should be developed and open-sourced. 

From the perspective of the methodology of developing AIGC detection algorithms, the existing research studies rely on the whole text to identify whether students are cheating or not. If solutions are developed on the whole text representations, students can alter the text with different evasion techniques discussed above. However, suppose solutions are developed with the main gist of the text, which contains the main topics or keywords and the summary. In that case, the solution may tolerate different evasion techniques. 

From the perspective of model results, most of the research implementations have been done on outputting the probability of AIGC on the whole text. Research on detecting AIGC probability paragraph-wise or sentence-wise would give justifiable predictions. The AIGC detection models should be able to generate the probability of AI-generated in different parts of the text. Additionally, there seems to be a gap in adding explainability to the prediction of models. Further research can be done by employing post hoc (after training and prediction) model-agnostic techniques like SHAP to approximate each word's attribution to the final prediction. This has been done to explain the predictions of text classification and sentiment analysis models \cite{mosca2022shap} and can be done to explain the predictions of AIGC detection algorithms.

From the perspective of the overall solution to fight academic misconduct, the major gap is that most of the research studies to solve this problem have been focused on simply classifying AI-generated or human-written and plagiarized or not. Other alternatives to solve the problem, such as preventing academic cheating, remodelling assessment strategies, and promoting ethical use of AI in academia, have not been focused compared to technical solutions like plagiarism and AIGC detection. 

\section{Discussion}\label{discussion}
It is evident from the experimentation performed and analysis from the survey of existing solutions in section 3 that a reliable solution to prevent academic misconduct due to LLMs is essential. This section of the paper discusses the possibility of solving the problem of AI plagiarism. Apart from the technical solutions, other alternative educational solutions that academic institutions should focus on to solve the problem have also been discussed. 

\subsection{Feasibility of AIGC Detection}
The survey was conducted on different aspects of text generation using recent LLMs, and the major question was whether it is possible to detect AI-generated text. Heikkiläarchive stated that it is extremely unlikely that there will be a tool that can detect AI-generated text with 100\% certainty in her article in MIT Technology review \cite{technologyreviewDetectingAIgenerated}. Muhammad Abdul-Mageed, a professor at the University of British Columbia, argues that it is difficult to do AIGC detection because the major objective of any LLMs is to generate more and more human-like text \cite{techwireasiaFindingReal}. The other main problem with AIGC detection is that if someone comes up with a great AIGC detection tool, there will be another LLM shortly on which it will fail. William H. Walters found that most AI detectors can distinguish papers generated from the GPT3.5 model and human-written papers but fail to identify papers generated from the recent GPT-4 model \cite{walters2023effectiveness}. Findings from Weber‑Wulf et al. \cite{weber2023testing} experiment on fooling 12 publicly available AIGC detectors and two commercial systems, Turnitin and PlgairismCheck, mostly used by universities by applying evasion techniques like translation and paraphrasing, suggested that a reliable solution for AIGC detection does not exist and may not exist.

\subsection{Alternative Educational Solutions}
After analysis of the preventive measures for academic cheating with more than 50 examples, including experience from various academic institutions, Bylieva et al. \cite{bylieva2020academic} concluded that the problem of academic cheating can not be solved only with pure technical measures because such technical war can last forever. Thus, academic institutions should consider and adopt non-technical alternative educational solutions. One of the major steps that should be considered is generating awareness of academic misconducts like plagiarism and improper use of AI tools like ChatGPT. In a survey conducted with 3405 students at an Australian university, it was reported that only 50\% of the students had read the plagiairism policy, and half of the students did not know what behaviour constitutes plagiairism\cite{gullifer2014has}. The office in a university that deals with complaints stated that most complaints were about the university failing to warn the students about plagiarism and its consequences \cite{theguardianUniversitiesNeed}. European Network for Academic Integrity (ENAI) also presented recommendations for academic institutions to promote the ethical use of AI tools like ChatGPT in academia. The recommendations include training both students and teachers about the ethical use of AI and the limitations of the bias and accuracy existing in such AI tools. Universities should also clarify the circumstances of using AI tools properly \cite{foltynek2023enai}. Thus, proper awareness of academic misconduct is vital.


Another solution to academic misconduct is to incorporate AI ethically and creatively in academic settings. Dr Amin Davodi expresses that students should be taught to use AI tools to learn, and there will be less chance of using AI for cheating \cite{elliiPreventStudents}. Universities should allow using tools like ChatGPT, but only up to a certain threshold. Students can use such tools for reviewing and correcting but cannot entirely copy-paste the content generated by those tools. This can be facilitated by setting up a threshold for ChatGPT-generated content in student's assignments. ChatGPT has been listed as a coauthor in research papers published \cite{stokel2023ChatGPT}. Universities and scientific publishers can create a threshold value for permitting the correct use of tools like ChatGPT. Jo Ann Oravec \cite{oravec2023artificial} proposes to design assignments so that ChatGPT can be transparent. The assignments can include the steps to compare the student's solution with ChatGPT, review and rewrite the assignment \cite{oravec2023artificial}. This way, tools like ChatGPT can be used more constructively. As recommended by ENAI, appropriate citation of the use of AI tools should be done for the ethical usage of AI \cite{foltynek2023enai}. Abd-Elaal et al. \cite{abd2019artificial} concluded that academic institutions should improve their plagiarism and fabrication policies. Likewise, ENAI also recommends that national guidance and institutional-level policies be developed concerning AI's ethical use in academia \cite{foltynek2023enai}.

Another solution for preventing academic misconduct is the improvement of current assessment strategies in universities. The assessment should focus more on assessing student's critical thinking and problem-solving skills rather than memorizing skills. Simple factual questions can be directly solved by using LLMs. Thus, students should be assigned creative projects with open-book exam questions that require critical thinking and problem-solving skills \cite{egan2018improving}. Another strategy that lecturers can use to asses students is by giving them visual or interactive questions, for instance, instructing a student to watch a video and answer questions based on the events in that video. This way, students cannot cheat on ChatGPT. Likewise, instructing students to give oral presentations or assessing students through viva examinations are some strategies that can prevent cheating \cite{harper2021detecting}. Another efficient trick to preparing assessments to prevent cheating is to prepare timed assignments \cite{holden2021academic}. With timed assignments, students cannot answer all questions by getting answers from ChatGPT in a limited time \cite{holden2021academic}.

\section{Conclusion}
In this research work, different ways by which academic misconduct is done and can be prevented were discussed. Specifically, the change in the landscape of academic cheating after introducing LLMs was discussed. We presented how the introduction of cheating with generative AI tools has changed other cheating forms like plagiarism. We surveyed both sides, i.e., the development of algorithms and tools to prevent AI-generated text and plagiarism and the evasion techniques applied to fool such algorithms and tools. With such evasion techniques and a combination of different tools, we demonstrated that existing solutions for plagiarism detection and AI-generated detection can not be relied upon. 

The survey identified several gaps in the existing datasets and solutions for plagiarism detection and AIGC detection. 
Similarly, considering the longevity of the technical war between AIGC detection and fooling AIGC detectors, we proposed other alternative educational solutions that academic institutions should use. Drawing upon the current scenario caused by LLMs in academia, the issue looks pretty concerning, and hence, policies and regulations regarding the ethical use of such tools in academia should be developed. 

Several future works can be done to improve the field of AIGC detection. Regardless of the amount of research done to solve the problem, there is a big question about the full reliability of existing AIGC detection tools and solutions. This is partly because there is no existing quality benchmarking for the problem. This can be partly solved by developing a benchmark AIGC detection quality dataset. Other experiments can be performed to solve the problem by applying a mixture of different methods to build more accurate AIGC detection algorithms. Apart from technical solutions, experiments on non-technical solutions can also be performed to solve the problem of AI-based plagiarism. Additionally, research on the explainability of the outputs of different AIGC detection models can be done to build trustworthy solutions. Thus, 
developing and releasing a quality benchmark dataset for the task, developing AIGC detection models with a mixture of different models, and adding explainability to the predictions coming from AIGC detection models are some of the major future works that can done.

\section*{Acknowledgments}
This was supported by the SFI Centre for Research Training in Machine Learning. 

\bibliographystyle{unsrt}  
\bibliography{references}

\end{document}